\documentstyle[aps]{revtex}
\newcommand{\blankline}{\vskip .3cm}
\newcommand{\blankblock}{\vskip 3cm}
\begin{document}
\twocolumn
\begin{titlepage}

\begin{flushright}
UM-P-95/24 \\
gr-qc/9506001 \\
June 1995 \\
\end{flushright}

\blankblock
\centerline{\LARGE The Quest for Quantum Gravity}
\blankline
\blankline
\blankline
\centerline{Gary K. Au\footnote{gau@tauon.ph.unimelb.edu.au.}}
\centerline{Research Centre for High Energy Physics}
\centerline{School of Physics}
\centerline{The University of Melbourne}
\centerline{Parkville 3052}
\centerline{Victoria}
\centerline{Australia}
\blankline

\begin{abstract}
One of the greatest challenges facing theoretical physics lies in reconciling
Einstein's classical theory of gravity - general relativity - with quantum
field theory. Although both theories have been experimentally supported in
their respective regimes, they seem mutually incompatible. This article
summarises the current status of the superstring approach to the problem,
the status of the Ashtekar program, and addresses the problem of time in
quantum gravity. It contains interviews with Abhay Ashtekar, Chris Isham,
and Edward Witten.
\end{abstract}
\vfill
\end{titlepage}

\newpage

\begin{center}
``It is very important that we do not all follow the same fashion... It's
necessary to increase the amount of variety... and the only way to do this
is to implore you few guys to take a risk with your lives that you will not
be heard of again, and go off in the wild blue yonder to see if you can
figure it out."
{\bf Richard Feynman}(1965), Nobel prize in physics award address.
\end{center}

\vspace{3mm}
{\Large{\bf Introduction}}
\vspace{3mm}

The road to quantum gravity (QG) has had a long and winding history.
Although we are unlikely to see the day of a
direct test of quantum gravity, such a theory is needed to describe the early
stages of the Big Bang. Around this time the universe would be about the size
of the Planck volume. Quantum gravity effects, ignored in particle physics
because of their weakness, would have a major influence on the subsequent
evolution of the universe. Also a QG theory is desperately needed to make the
standard model a consistent physical framework.

Why are we spending so much time on such a lofty goal?
Richard Feynman commented on the aims of science:
``If you expect science to give you all the answers to the wonderful
questions about what we are, where we are going, and what the meaning of the
universe is, then I think you could easily become disillusioned and look for a
mystic answer to these problems.

``The way I think of what we're doing is, we're exploring - we're trying to
find
out as much as we can about the world. People say to me, `Are you looking for
the ultimate laws of physics?' No, I'm not. If it turns out there is a
simple, ultimate law which explains everything, so be it; that would be very
nice to discover. If it turns out it's like an onion, with millions of layers,
and we're sick and tired of looking at the layers, then that's the way it is.
But whatever way it comes out, it's nature, and she's going to come out the
way she is! Therefore when we go to investigate it we shouldn't predecide what
it is we're going to find, except to try and find out more.\footnote[1]{No
Ordinary
Genius - The Illustrated Richard Feynman, Christopher Sykes, W.W.Norton and
Company,
1994, p251.}"

\vspace{3mm}
{\Large{\bf The Problem of Time}}
\vspace{3mm}

A key question which faces any approach to QG is the `problem of time': how
does a time variable, with its special properties in relation to the
probability interpretation of quantum mechanics (QM), emerge? A related topic
which
has been attracting increasing attention, especially in the context of
`quantum cosmology' is the `consistent-histories' approach to quantum theory.
One of the most exciting features of the programme is the possibility of
changing quantum theory itself in the context of the physics of the early
Universe and the way in which it came into being.

The basic problem lies in the incompatible way in which the concept of time is
treated in QM and general relativity (GR). One of the main researchers studying
this deep problem is Chris Isham at Imperial College, London.

In QM time evolution is described by the Schr\"odinger equation. However the
time parameter that comes into the equation is a background parameter. This is
unlike other quantities which can be represented by Hermitian operators. So
from the very beginning QM attaches a special significance to the idea of
time. It's like the steady ticking of an external clock hand in the
background, which we have no control over. Can we physically measure this
quantity? Is it possible to reformulate QM so that the idea of a background
time doesn't appear? According to Chris: ``It is possible to do this in a
formal sort of way but it does not really get you far. The basic fact is that
conventional quantum theory presupposes an external time whose
ontological\footnote[2]{Ontology is the part of metaphysics concerned with the
nature of existence.} status is the same as it possesses in classical
Newtonian physics. In relativistic quantum theories this is generalised to
become a background Minkowskian spacetime - so it is the causal structure that
is fixed. There are some intriguing mathematical theorems (going back to Pauli
I believe) showing that you cannot construct a {\em quantum} clock that would
exactly measure this background time. In that sense, some people might want to
consider `time' as part of the external classical realm that Bohr postulated
was necessary to interpret the equations of quantum theory. I have always
suspected that Bohr himself would have regarded the subject of `quantum
gravity' as a non starter!"

Why does time deserve a special status in QM? Well, if you could raise it to
the status of an operator $T$ then its conjugate variable would be the
energy $H$. These would need satisfy the commutation relations $[T,H]=i\hbar$.
Mathematically it can be shown that it is impossible to have a pair of
Hermitian
operators satisfying these conditions, in which one of them has only positive
eigenvalues. This relates back to Chris' remark about the problem of quantum
clocks. In other words, we have the problem of a system with negative energy.

In QM, measurements of observables are taken at fixed times, a complete set of
commuting operators (defining the maximum information you can obtain from a
system) is required at fixed times, and the scalar product on the Hilbert
space of states has to be conserved under time evolution of the Schr\"odinger
equation. Time plays a foundational role in the technical and conceptual
aspects of quantum theory. However it is an idea grounded in the world of
continuum
mathematics. If it breaks down at the Planck scale, where some believe a
discrete
structure will emerge, what are we left with? Indeed, how much of QM will hold?

According to John Wheeler: ``The word `time' was not handed down from heaven as
a gift from on high; the idea of time is a word invented by man, and if it has
puzzlements connected with it, whose fault is it? It's our fault for having
invented and used the word.\footnote[3]{Stephen Hawking's A Brief History of
Time: A Reader's Companion, prepared by Gene Stone, Bantam Books, 1992,
p125.}" Taking this into mind, could we redefine time so that it reduces to
the familiar clock ticking concept in some classical limit?

Chris Isham is very sympathetic to this view: ``I would bet strongly
that whatever the `final' theory of quantum gravity will be, we shall see our
standard notion of time only emerging in some semi-classical sense."

The role of time in relativity is very different to its role in QM. Relativity
treats time on equal par with the other three spatial variables. Indeed, time
and space are unified into a four dimensional spacetime manifold.

On this matter Chris explains: ``The notion of a single, 4-dimensional,
`spacetime'
(rather than separated 3-dimensional space plus 1-dimensional time) first
appears in special relativity (SR) because of Einstein's realisation that any
absolute split of 3+1 is observer-dependent. General relativity works with a
curved version of the flat space-time of special relativity.

``In both special and general relativity a `moment of time' corresponds to a
single space-like hypersurface in the space-time. Time itself appears as the
parameter that labels the elements of a one-parameter foliation\footnote[4]{In
SR, a
foliation is a one-parameter family of three-dimensional {\em flat} surfaces
(hyperplanes) which, taken all together, fill out the entire spacetime. In GR,
these
surfaces are curved.}
of space-time by such surfaces. In
the case of special relativity you restrict your attention to the case where
the space-like surfaces are (i) hyperplanes; and which (ii) are mapped into
each other by actions of the Poincar\'{e} group. The main challenge in the
quantum theory is to show that all such admissible families give the same
physical answers.

``The situation in general relativity is more complex. One says that a
hypersurface is {\em space-like} if the vectors tangent to each point of the
surface are space-like. The idea here is that the Lorentzian metric on
spacetime induces on each tangent space a copy of the Minkowski metric of
special relativity, and it is with respect to this latter metric that the
tangent vector has to be space-like. Put slightly less rigourously, a
hypersurface is space-like if any `infinitesimal' transformation in the surface
always points in a space-like direction."

The crucial point here is that the problems encountered when incorporating SR
into QM can
be overcome. When we try to add GR to the QM picture, the intricacies are much
more difficult to handle.

With a {\it curved} spacetime equipped with a Lorentzian metric tensor, how
many
ways are there to foliate spacetime as a one parameter family of space-like
surfaces? Chris continues: ``In some spacetimes there is a topological
obstruction
to performing {\em any} global foliations. But if such foliations are possible
then there
will be an uncountable number of them, each corresponding (in general
relativity) to an allowed definition of time." In other words, unlike in SR, we
cannot
define a unique time according to which we evolve a system. This
is the main difference between the concept of time in special and in
(classical) general relativity. It is an important aspect to the problem of
time in QG. On one hand we have a Hamiltonian formulation of QM for privileged
SR
space-like surfaces. On the other we have general covariance in GR which states
that no one set of space-like surfaces has privilege as a reference system
over others.

When we get to Planck scale physics, spacetime geometry could be subject
to Heisenberg's uncertainty principle, varying about quantum mechanically. In
other words the GR metric tensor $g_{\mu\nu}$ could have fluctuating
components! How
could one define a light cone and hence a space-like, light-like, or time-like
separation under such circumstances? Without such a definition, the idea of a
space-like surface seems invalid. Perhaps the idea of a spacetime (a continuous
manifold of points) doesn't have physical meaning in the QG regime, where the
light cone could be smeared out.

To this Chris says: ``You cannot define a light cone etc. except in some
`background' sense, and this perhaps {\em is} the essence of the problem of
time in QG: there is no fixed microcausal structure which can be used to
construct a relativistic quantum theory in any of the standard ways.

``However, this does {\em not} rule out a priori a continuous manifold:
the ideas of causal structure and continuous manifold are in no way
synonymous. You can easily have the latter without the former
(although I myself am very sympathetic to the idea that the idea of
a `continuum spacetime' is not something that will carry across to
the `final' QG theory)."

One possibility at the Planck scale is for topology changes in spacetime
geometry. Fixing a background topology and differential structure to that of
Minkowski space, as is done in quantum field theory, seems to presuppose a
lot at the Planck scale.

If we stick with a continuum formulation of QM based on complex numbers,
could we be missing out on revealing some sort of discrete structure
at the Planck scale? One can see how a discrete theory could reduce to
a continuum one in the large scale limit, but to shed light on a discrete
theory while working from the perspective of a continuum one seems
difficult to achieve.

Chris has felt this for a long time. One of the many reasons
why he is so interested in the new decoherent-histories approach to
quantum theory, and is developing his own quantum-logical version, is that the
role of
the continuum is isolated in a particularly efficient way. In his scheme it is
just the space in which the decoherence functionals take their
values. He says: ``It is sometimes suggested that near the Planck length a
more `combinatorial' approach to physics may be appropriate, and I
have some sympathy with this view. I think it would be much easier
to think about such schemes within this new approach to quantum
theory than in the old Hilbert space one."

Is superstring theory able to handle this fundamental problem of time in
its present perturbative formulation?

Chris replies: ``No. Although it is essential to emphasise that the `problem of
time' is not a single problem (or even collection of problems) that
applies willy-nilly to any attempt to construct a quantum theory of
gravity. It is very approach-dependent and looks very different in
different approaches to quantum gravity. For example, if space and
time themselves do `emerge' in some non-perturbative way from
superstring theory, then questions concerning the nature of time
will look very different from those that arise naturally in, for
example, the Wheeler-DeWitt approach to canonical quantum gravity.
That is one of the many reasons why it is important to keep
trying to find a new non-perturbative way of looking at superstring
theory."

The universe is modelled as a closed system. In its present formulation QM is
inadequate for handling such models. After all, how does one define a
time external to the universe? How successful have attempts been to define
an internal time using say, matter fields?

``They have been getting better. In a recent paper\footnote[6]{
Dust As A Standard Of Space And Time In Canonical Quantum Gravity.
By J.David Brown (North Carolina State U.), Karel V. Kuchar (Utah U.),
GRQC-9409001, Aug 1994. 57pp.
e-Print Archive: gr-qc@xxx.lanl.gov - 9409001.}, Karel Kuchar
(who is {\em the} world expert on this type of thing) has come extremely
close. Just how close depends precisely on what one decides one is
trying to achieve. From one perspective one could say that he {\em has}
succeeded in a technical sense, although he would be the first to
agree that his particular definition of time is not too compelling
in any fundamental physical sense."

Karel uses dust particles\footnote[7]{Dust is a basic type of matter which
only experiences gravity, and no other
interaction.} to fix points in space and identify moments in time. The dust
acts like a reference fluid analogous to the the old aether idea but with the
significant difference that the aether was usually regarded as fixed, i.e.
part of the background structure, whereas these dust particles do obey
equations of motion. Hence in a sense they are dynamic, as is the geometry
of spacetime.

One of the problems of working in QG is the lack of empirical data. One clue
comes from the 1992 COBE results. These showed variations in the blackbody
microwave background coming from different directions of the sky.
These date back to 300,000 years after the Big Bang and are consistent with
the inflationary scenario. The observations provide evidence for the idea of
quantum fluctuations as the origin of the galaxies. Could they have any
implication for the QG regime of $10^{-43}$ seconds after the Big Bang?

``Stephen Hawking has claimed that they support the ideas on the
very early universe put forward by himself and Jim Hartle. However, the
COBE data is currently being cited by a number of people in support of a
variety
of different views, and it is not clear yet what the outcome will be."

\vspace{3mm}
{\Large{\bf Consistent Histories}}
\vspace{3mm}

An approach which attempts to remove reference to external time in QM is
consistent histories, pioneered by Griffiths, Omn\`{e}s, Gell-Mann and Hartle.
It generalises QM via a sum over spacetime histories
formulation. The motivation is to describe the QM of closed systems in the
context of quantum cosmology. It hopes to shed light on the initial condition
of the universe. The basic idea in this scheme is a QM history.
The scheme therefore de-emphasises the notion of a QM event at a single moment
in time. It makes no reference to external observers, classical measurement
apparatus, or wave function collapse (reduction of the state vector).
However, it is possible to recover Bohr's formulation of QM under suitable
conditions. Chris Isham is working on a version of this idea.

``Most work on this approach has been applied to standard quantum theory, not
quantum gravity. To bring in quantum gravity Hartle has discussed the
possibility of using solutions of the Wheeler-DeWitt equation in the
context of Euclidean path integrals. I myself am working on this general
problem but from a completely different tack. I and my colleagues have been
studying consistent histories using a new type of `quantum logic' of
propositions about
generalised spacetime structures. But the general idea in
this formalism still holds: i.e. properties can only be ascribed to a
`history' when it is part of a consistent set. Typically this
requires deliberately loosing information by coarse-graining. One's
expectation is that `time' is something of this sort: i.e., it is really
only part of a semi-classical world that will `fade away' if one probes
too closely."

The scheme's main goal is to assign probabilities to families of histories in a
closed system. A history is defined to be a sequence of QM events at a
succession of times. QM events are tested for using projection operators which
satisfy the properties of exhaustivity and mutual exclusiveness. Two types of
projections are defined: fine-grained and coarse-grained ones.

The main problem in the formalism is interference. Interference between
different histories forbids assignment of probabilities since probabilities
have to satisfy certain axioms to be meaningful. They must be non-negative,
normalised, and they must sum to unity. Complex probability amplitudes
can be assigned to histories, but a probability is defined as the
modulus squared of an amplitude. This operation introduces cross terms
which violate the probability axioms.

To handle this problem `consistency conditions' are introduced. These
determine sets of histories which negligibly interfere, and to which
probabilities may be assigned. These conditions can be combined into a
decoherence functional, which is the main calculational tool of the QM of
history. It must satisfy the properties of hermiticity, positivity,
normalisation, and the superposition principle. The functional is computed
using standard quantum mechanics. Griffiths, Omn\`{e}s, Hartle and Gell-Mann's
contribution is the new probability interpretation. Here the notion of
`superposition' does not mean quite the same as it does in standard
way of superposing wave-functions. It is more a question that the
decoherence functional must be additive with respect to a disjoint
sum of histories.

In order to decouple interfering histories by coarse graining we require the
idea of decoherence. Consistent (decoherent) histories are then sets which
obey the consistency conditions. Consistency is a property of families of
histories, of which there may be many.

But what is the mechanism of decoherence? It has been postulated\footnote[8]{
Physics Today 44, 36, (1991), {\em Decoherence}, Wojicek Zurek.} that the
environment decoheres by acting like a sink, by draining off coherence -
similar to a piece of blotting paper soaking up excess ink. But if the system
is a
closed QM universe how do you define an environment? What other methods could
there be to decohere closed systems?

To this problem Chris remarks: ``The mechanism for decoherence in a truly
closed
universe is, to my mind, still rather problematic. There have been attempts to,
for
example, identify part of the gravitational-field complex as the
environment, and part as `really quantum' freedom in this
environment. But I am doubtful whether this is really a universal
notion. The examples I have seen have all been for very simplified models."

In a closed QM universe, would it be possible to use black holes as decoherence
devices? When matter goes into a black hole, the only properties of the hole
which are
observable via its distant electromagnetic and gravity fields are mass,
charge, and angular momentum. All other information is lost. Would this
loss include `interference' information between histories?

Unfortunately Chris says: ``One could do that when studying quantum theory
in a {\em fixed}
background, since then one could consider a background containing
black holes. However, this is not too meaningful a priori in a full
theory of quantum gravity itself since the black holes will also
have to be quantised in some sense."

Jim Hartle made the important observation that one can talk about
`generalised histories' in QG in which the word `history' need {\em not}
presuppose a conventional type of time parameter. According to Chris:
``The basic entity in the scheme is a `possible universe'. A
simple, but informative example would be a Lorentzian geometry that is
{\em not} globally hyperbolic. Such a spacetime does not admit a global
foliation by space-like surfaces, and hence there is no global time
parameter. Nevertheless, it is a viable classical possibility since time
still makes sense locally (i.e. the local proper time along a world line).
This was one of the original motivating examples suggested by Hartle.

``However, my own interest in this scheme is when the basic entity is
something more basic that does not necessarily include any manifest
time-like features at all. This would be in line with the general
idea that time appears only in some semi-classical limit. Of course,
this leaves quite open what types of object one would actually use
as `possible universes' since this depends entirely on the actual
theory that is to be developed (the decoherent-histories approach is
not a {\em theory} per se, it is more a {\em theoretical framework}
in which new types of theory can be developed). For example, I
am currently thinking about the possibility of constructing such a
theory in which the basic entities are simply point-set topologies.

``In theories of this generalised type Hilbert spaces do
not appear as basic mathematical structures in the same way that
they do in normal quantum theory; indeed, in the scheme that I am
developing the basic mathematical structure is much closer to that
of quantum logic (i.e. an algebraic approach) than it is to Hilbert
space methodology. On the other hand, in any theory in which time
could be shown to `emerge' in some coarse-grained limit I would
expect the usual Hilbert space structure to emerge too at the same
level. However, as no one has yet written down an explicit example
of this type, it is hard to be more specific!"

\vspace{3mm}
{\Large{\bf The Ashtekar Program}}
\vspace{3mm}

People have approached the QG problem from many different directions, some
more successful than others. Most attempts
suffer from seemingly insurmountable problems. A short list of avenues
includes higher derivative Lagrangians, twistors,
induced gravity, Kaluza-Klein theories, Euclidean quantum gravity, covariant
perturbation theory, discrete gravity, non-linear quantum mechanics, spin
networks, asymptotic quantisation, quantum cosmology, the decoherent histories
just covered, quantum field theory in curved spacetime, superstrings,
and canonical gravity. The Ashtekar program belongs to the last category.

Abhay Ashtekar is at the Centre for Gravitational Physics and Geometry,
Pennsylvania. Although his program
involves esoteric mathematics, some of its key features can be described
qualitatively.

An obvious question that people might ask about QG research is: Why are we
spending so much time and effort trying to obtain a theory which is most
probably
untestable, and which won't have any practical application?
Abhay doesn't quite agree with the assumptions here. He says: ``We are seeking
a {\it physical} theory and therefore it will definitely have testable
predictions. The initial tests will be probably indirect. For example,
we have this great puzzle for over 50 years: quantum field theory
provides incredibly accurate predictions in particle physics yet it
is, in a sense, only a set of calculational recipes, without a
coherent, complete mathematical framework. Again, the recipes work
incredibly well and should be taken seriously. But surely, they are
incomplete in some essential way. The ultraviolet divergences arise
largely because we assume that space-time is a continuum at all
scales. If a quantum theory of gravity provides an alternate picture
of space-time, it should lead to an alternate way of doing quantum
field theories which is coherent and mathematically consistent. So, in
a sense, the vast experimental data we have accumulated in particle
physics could provide tests of the space-time models that come from
quantum gravity.

``Furthermore, it is possible that quantum gravity will also lead to
`practical applications'. Recall that, when Einstein discovered
special relativity, most people thought that it was an esoteric theory
which would have little impact, if any, on daily life. However, the
theory then led to $E = mc^2$ which has had profound ramifications on
the entire world-order of this century. Quantum gravity will also
change our understanding of space and time radically and may therefore
have some really deep practical implications. I would hope that they
would be more peaceful in nature!

``But of course we don't have the faintest clue of such applications
today and, as you indicate, it is beyond today's technology to test
the quantum gravity effects {\it directly}. The primary motivation is
really conceptual. Again, there is a similarity with special
relativity. Just as the primary motivation for special relativity came
from the incompatibility between Newtonian mechanics and Maxwell's
electrodynamics, the two pillars of the 19th century physics, the
primary motivation from quantum gravity comes from the tension between
general relativity and quantum mechanics. There should be a deeper
theory which unifies the principles of both in such a way that these
two theories emerge in suitable approximations. That is the theory
that workers in quantum gravity are trying to construct."

Abhay's program was inspired by the failure to find a
renormalisable perturbative theory of quantum GR (QGR). What this is, and what
would have happened if this failed program had been successful?

He explains in depth: ``Perturbation theory assumes that answers to
physical questions can
be obtained by a method of successive approximations, generally
expressed in terms of a power series in the coupling constant. We
routinely use such approximations already in quantum mechanics. In
quantum field theory, however, there is a key difference. Whereas in
quantum mechanics, we know from general principles that the exact
answer exists and is finite and perturbative techniques are used only
as a computational tool, in realistic quantum field theories in
4-dimensions, we do not have an assurance that there is an underlying
well-defined theory; the perturbation series itself is being used to
define the theory. To make matters worse, each term in the
perturbative expansion diverges because one is allowing virtual
processes of arbitrarily large momenta. In renormalisable theories,
the infinities that arise are of a special nature; they can be
absorbed into a finite number of parameters associated with the theory
such as charges, masses and coupling constants. Once these parameters
are renormalized by the appropriate infinite factors, all individual
terms in the perturbative expansion become finite. The actual
renormalized parameters can be determined by a finite number of
experiments. Further experiments then provide tests of the theory. In
a non-renormalisable theory, this can not be achieved and therefore
the perturbation expansion has no predictive power.

``Another way of stating the difference is that, in a renormalisable
theory, predictions for phenomena at a given length scale are
insensitive to what is happening at much smaller scales. So, for
example, since QED is renormalisable, its predictions for processes
at, say, a GeV scale are insensitive to all the goings on at the Planck
scale ($10^{19}$ GeV). In non-renormalisable theories, this is not the
case; the predictions for one length scale {\it are} sensitive to what
is happening at much smaller scales. It is thus harder to extract the
physical content.

``If perturbative general relativity had turned out to be
renormalisable, on one hand, life would have been simple. One could
have calculated scattering cross-sections in gravity using familiar
methods. On the conceptual side, however, at least some people
including me, would have been disappointed. It would have meant that
most of physics is insensitive to the `true' small scale structure
of space-time and hence it would have been harder to `probe' this
structure. I have expressed my unease with the current status of
quantum field theory. It is the non-renormalizability of gravity that
forces us to seek the true micro-structure of space-time and holds
clues for obtaining something which goes beyond (the calculational
recipes of) quantum field theory."

Perturbation theories assume spacetime is a continuum at all
scales. Could this be a reason why they fail for QG, where the short
distance behaviour of spacetime might not be a continuum? An analogy
has been made that spacetime is a `foam' of sorts at Planck scales,
subject to Heisenberg's uncertainty principle.

``That's right! The key infinity in perturbative treatments is the
ultraviolet one where processes involve arbitrarily large energies.
Surely, when the energy becomes bigger, gravity should become more and
more important, and the space-time geometry, more and more
non-trivial.  Yet, in the perturbative treatments, one uses a fixed,
continuum background space-time. Because space-time geometry is a
dynamical entity in general relativity, it seems clear that in quantum
gravity we can not {\it presuppose} what the micro-structure of
geometry must be; we should let the theory itself tell us. Of course,
one might have been lucky and made the right guess at the start. But
the failure of perturbative methods tells us that this was not the
case with the continuum hypothesis."

Might quantum GR exist non-perturbatively, i.e. as an exact
theory?

``Yes, it might. We don't have any clear evidence to the contrary.
Incidentally, there do exist theories in 3 space-time dimensions which
are non-renormalisable but exactly soluble. Note that I am {\it not}
saying that quantum GR must exist non-perturbatively. However, since
there are qualitative reasons to expect that the non-perturbative
theory will be very different from the perturbative one, the question
is well worth investigating. More importantly, even if the answer
turns out to be in the negative, the exercise will have been extremely
worthwhile because it has already provided many new mathematical and
conceptual tools to construct quantum field theories without a
background space-time. Like many others, I firmly believe that, at a
fundamental level, the final solution will have to be background
independent; space-time will emerge only as an approximation. And we
have very little experience with quantum field theories which have an
infinite number of degrees of freedom and which do not use a background
space-time."

Do you see the need for a fundamental revision of the present
concepts on which the standard model is based? In other words, are we
in need of a conceptual revolution to solve this problem? I once spoke
to Bill Unruh, and he
didn't think QG could be obtained through `tinkering with
mathematics'. He said we needed to feed in new conceptual ideas.

According to Roger Penrose ``...if there is to be a final theory, it
could only be a scheme of a very different nature. Rather than being a
physical theory in the ordinary sense, it would have to be a principle
- a mathematical principle whose implementation might itself involve
non-mechanical subtlety." Is this your point of view as well?

``I believe that, at a fundamental level, the continuum picture has
to go. This would be a profound change, both in terms of physics and
mathematics since the picture is embedded so deeply in the conceptual
fabric of physics. So, I agree with Bill that `tinkering' with
mathematics alone will not suffice. However, I also feel that new
mathematics {\it will} be needed. (I mean mathematics that was not
previously used in physics.)  This has happened with most big
breakthroughs in physics. Newtonian physics needed calculus; general
relativity, differential geometry and quantum mechanics, Hilbert
spaces and operator algebras. Indeed, without access to new
mathematics, it would not have been possible to formulate new
questions, let alone analyse them! Certainly, what Roger suggests is a
possibility."

Most approaches to the QG problem seem to take QM at face value,
and try to modify GR in some way, or take GR as a weak field
limit. This is the case for superstring theory, where GR appears as a
low energy limit. The dimensional nature of the basic Planck units -
the Planck length, time, and energy - lends credence to the idea that
a QG theory could reproduce GR in regimes well away from the Planck
scale. If QM is only an approximate theory itself, which fails to hold
at Planck scales, then in the spirit of Roger Penrose's suggestion, QM will
also have
to undergo a revision. Abhay feels that this is in part why {\it all}
attempts at quantum gravity that are being pursued are likely to be incomplete.
``However", he says, ``it is
very difficult to provide concrete models of how QM should be revised.
I think the best bet is to push all viable approaches to the end and
if cracks develop, they will suggest how we should change QM."

Loop variables gravity seems like a conservative approach compared
to superstring theory. One major difference is that according to your
program, quantum gravity can be obtained without a unification of all
forces. In the superstring theory, such a unification is vital for
consistency. To be consistent, superstrings also need supersymmetry (a
mathematical relationship between bosons and fermions) and higher
spacetime dimensions (nine space and one time). The extra six spatial
dimensions are supposed to be compactified into a very small
region. Do you see these differences as complementary or conflicting
issues?
``On the whole, I think the differences are complementary.
String theory provides powerful constraints on allowable
interactions.  On the other hand, it has proved difficult to analyse
string theory non-perturbatively, without assuming a background
space-time. As I indicated above, our approach provides concepts and
tools for a background independent approach. Over the last year or so,
more concrete calculations have been proposed which may pull the two
approaches closer.

``I should also point out that we don't have a proof that quantum GR
exists non-perturbatively. However, the tools required to analyse this
issue finally exist. It is possible that we will find, e.g., that
supersymmetry or some specific couplings are necessary. This year is
an exciting one for our program because we have begun to analyse such
issues."

I spoke to Peter Bergmann (one of the founders of the
geometrodynamical route to QG, and a co-worker of Einstein) while I
was an undergraduate and he was very skeptical of superstrings. Is it
the same with you, considering most of the work in this field has been
grounded in perturbation theory?

``I am perhaps more impressed than Peter by the mathematical
successes of string theory. I agree that the use of a background
space-time is a severe weakness. However, the very recent results due
to Ashoke Sen\footnote[9]{See for instance, {\em Extremal Black Holes and
Elementary String States}, Ashoke Sen (ICTP, Trieste and Tata Inst.).
TIFR-TH-95-19,
Apr 1995. 16pp. e-Print Archive: hep-th/9504147.}
and others on black holes within string theory are
providing some glimpses of the possible relation between perturbative
results and non-trivial geometries."

Unlike the other forces, the gravitational force is non-linear. In
contrast, the electrical forces are linear because the total force due
to a number of charges is just the sum of the forces due to each
charge acting alone. This situation corresponds to the vector law of
addition. Now in the gravity case, we have a non-linear combination of
forces. This is because all forms of energy act
as a source of gravity and inertia. This is true for the gravitational
potential energy existing between two separated masses.
Thus the total gravitational force involved when
two objects interact does not follow the usual linear vector addition
law. Is this non-linearity one of the root causes of all the problems
in quantising gravity?

``This profound non-linearity is part of the difficulty. But this
part is technical. The the most important difficulty, in my view, is
conceptual: it lies in the fact that gravity is encoded in the very
geometry of space-time. To quantise gravity, one has to quantise
geometry. One has to learn to do physics and mathematics in absence of
space-time."

I guess one of the things readers want to know is how everything
reduces down to the picture of an apple falling from a tree. Newton's
idea of a `force' pulling the apple down is intuitive. In GR, this
picture was replaced by the concept of the curvature of a four
dimensional spacetime continuum. The apple's mass curves spacetime,
and the Earth's mass curves spacetime. This curvature tells how the
masses should move. The two masses feel the combined curvature, and
fall towards each. In QFT, this `attractive' picture is replaced by
the concept of the exchange of particles called gravitons. The
gravitons mediate the gravitational pull. Does the loop variables
theory preserve this idea?

``Yes. Although in our approach gravitons don't exist at a
fundamental level, in the low energy approximation one can identify
certain loop states with gravitons and then the picture is the one you
indicate.

``Incidentally, the loop variables also have a direct experimental
significance. If you move an electron around a closed loop in a
gravitational field, its spin undergoes a rotation. The basic loop
variable just measures this rotation. Furthermore, if you know the
rotation for all closed loops, you know the gravitational field
completely."

One of the problems with gravitons is that they seem to make sense
only when the gravitational field of GR is linearised, i.e. split up
into a flat space metric, and a perturbation. Gravitons are like
ripples in a flat spacetime background. Conventional particle theory
tries to quantise these undulations, while leaving the background classical.
When the background starts curving noticeably itself, what reference is
there for these ripples? The problem seems to be that the arena and performers
can't be distinguished anymore.

``Gravitons are indeed ripples on a space-time background. In
the perturbative approach, one does quantise these undulations leaving
the background untouched. In our approach, the situation is different.
Very roughly speaking, Minkowski space by itself is represented as a
quantum loop state - which we call a weave - and certain nearby states
- embroidered weaves - can be thought of as gravitons propagating in
Minkowski space. However, as I indicated above, the picture works only
for low energy gravitons. In this regime, one can distinguish the
arena from the performers."

\vspace{3mm}
{\Large{\bf Geometrodynamics}}
\vspace{3mm}

Abhay's program takes conventional GR coupled to matter as the leaping point.
It is based on the `canonical' approach to QG whereby the full four dimensional
spacetime is split up into an observer dependent $3+1$. The first step in
any canonical approach is the Hamiltonian formulation of the theory.
This was achieved in the early sixties by Dirac, Bergmann, Arnowitt,
Deser, Misner and others. In their formulation, the basic variable is
the metric on the three dimensional space; it captures the intrinsic
geometry of the hypersurface. The Hamiltonian framework was therefore
baptised `geometrodynamics' by Wheeler. Unfortunately, in this
framework, the equations of the theory are complicated and therefore,
as far as the full theory is concerned, the quantisation program
itself did not really take off.

In geometrodynamics the canonical variables are the intrinsic geometry and the
extrinsic curvature
of the 3D spatial hypersurface (which is embedded in a 4D Riemannian
spacetime).
The subject has close links with Gauss' Theorem Egregium.

Abhay explains: ``The intrinsic geometry of a 3-dimensional hypersurface is the
geometry of space that we can `experience directly'. It tells us,
for example, how to measure lengths of curves lying in the surface.
The extrinsic curvature on the other hand, tells us how the surface is
embedded in the 4-dimensional space-time. To see the difference, let
us go one dimension down and consider a cylinder embedded in a
3-dimensional Euclidean space. Since the cylinder can be obtained by
rolling up a piece of paper, its intrinsic geometry is the same as
that of the paper, i.e., a 2-dimensional plane. The difference lies in
how the paper and the cylinder sit in 3-dimensional
space. Technically, their extrinsic curvature is different.

``Coming back to GR, the extrinsic curvature measures how the intrinsic
geometry is changing in time. So, if we think of the intrinsic geometry
as the configuration variable of general relativity, as one does in
geometrodynamics, then extrinsic curvature can be thought of as the
conjugate momentum. In quantum geometrodynamics, one would subject the
two to the Heisenberg uncertainty relations."

By splitting spacetime into (3+1)D, are you treating the spatial
and time co-ordinates on a different footing? Physically, this would
correspond to choosing a special observer, whose wristwatch keeps
track of the distortions in the 3D spatial hypersurface. This method
is very different to the `covariant' schemes of quantisation like the
path integral approach.

``That is quite right. From aesthetic considerations, it would be
nice to use, say, path integrals. However, no one has seen a way to
make mathematical sense of path integrals in GR beyond certain simple
minisuperspaces (which result when you freeze all but a finite number
of degrees of freedom of the gravitational field and replace a quantum
field theory problem by a quantum mechanics problem). The
mathematical framework that my colleagues and I have developed over
the last two years is, in principle, also applicable to the Euclidean
path integral approach. However, there are two key difficulties. The
first is conceptual. Even if one did have a complete Euclidean theory,
it seems very difficult to extract physical predictions from it since,
in quantum gravity, there is no analog of the `Wick rotation' which
replaces the time coordinate $t$ by $it$ in Minkowskian quantum field
theories thereby enabling one to pass from the Euclidean to the
Lorentzian regime (i.e. from Schwinger functions to the Wightman
functions). The second difficulty is technical, associated with the
specific form of Einstein's action."

One of the main criticisms of the approach is that it breaks the
general covariance of GR. Would the general covariance concept have
any meaning if the continuum picture of spacetime is lost? One of the
ideas brought up by your program was the possibility of spacetime being
combinatoric instead of geometric in nature, at the Planck scale.

``For technical clarity, I should first point out that, while
canonical approaches do lack {\it manifest} covariance, they are,
nonetheless covariant; for example, the full Poincar\'{e} group has a
well-defined action on the phase space of electromagnetic fields in
Minkowski space. With this side remark out out the way, let me answer
your question. If anything resembling our approach is correct, then,
in the quantum theory, there will be no such thing as a space-time
geometry at a fundamental level. The picture, as you say, will be
combinatorial and therefore the diffeomorphism group will not have a
preferred role. Space-times will arise only semi-classically. In this
approximation, one can ask if there is 4-dimensional diffeomorphism
invariance. The answer will be in the affirmative. However, it will be
arrived at from the Hamiltonian picture; the 4-dimensional covariance
will not be manifest. This is a drawback of the approach but only
from an aesthetic viewpoint.

``Incidentally, for these reasons, I had first tried to use the
covariant phase space for general relativity and avoid the 3+1
splitting. That approach faced technical problems and I now believe
that, even if the technical problems could be resolved, it would not
lead to an interesting theory. I won't be surprised, however, if there
is another way to ensure manifest covariance in the semi-classical
regime."

By casting GR into Hamiltonian form, it is found that the theory is a
dynamically constrained system. These restrictions go by the name of
the Hamiltonian and diffeomorphism constraints. What physical significance
do they have?

``Constraints of this type signal the existence of a large gauge
group. In electrodynamics, for example, there is an analogous
constraint, the Gauss law, which says that the electric fields should
be divergence-free. It ensures that the classical (and quantum)
theory is gauge invariant. Higgs pointed out, quite early in the game
that the diffeomorphism constraint of general relativity is rather
similar. It ensures that the spatial diffeomorphisms - active
coordinate transformations on the 3-dimensional hypersurface - don't
change physics. This constraint is universal in the sense that its
form and meaning is the same for all gravity theories in which there
is no background metric.

``The Hamiltonian constraint is more tricky. It is responsible for time
evolution, i.e. dynamics. If you treat it in the same fashion as the
diffeomorphism constraint, one is led to what people call `the frozen [time]
formalism' in which nothing evolves and physical states are
represented by entire space-times. There is nothing wrong with this
picture. It is like doing Hamiltonian dynamics in terms of constants
of motion; it is therefore inconvenient from from technical and
interpretational perspectives."

I have heard that the constraint equations are (high order)
polynomials in terms of the basic canonical geometrodynamical
variables - the 3-metrics and extrinsic curvatures. What effect does
this have on trying to quantise the theory?

``It simplifies the technical problems of quantisation enormously.
In quantum field theories, what one might naively think of as
operators are really operator valued distributions. Just as products
of the Dirac delta distributions are ill-defined, products of these
`operators' are also ill-defined. This is the origin of the
infinities. So, if you have an expression which is non-polynomial in
your basic operators, in general, there is very little chance that you
would be able to regulate it and extract a meaningful operator. If the
expression is a {\it low} order polynomial, the chances of success are much
greater. At various levels of rigour, this simplicity has been
exploited by Rovelli, Smolin, Br\"{u}gmann, Gambini, Pullin, Morales,
Nicolai and Matschull to solve the quantum constraints. No such
solutions exist in quantum geometrodynamics.

``A simplification came in the mid eighties when it was realized that
there is another way of obtaining the Hamiltonian formulation, where
the basic variable is a connection which enables one to parallel
transport chiral fermions. Thus, general relativity can also be
regarded as connection-dynamics. This formulation has two key
advantages. The first is conceptual: general relativity is brought closer
to gauge theories which govern all other basic forces of Nature. The
second simplification is technical: equations of the theory become low
order polynomials."

\vspace{3mm}
{\Large{\bf The Ashtekar Variables}}
\vspace{3mm}

At this stage, you came up with the idea of the Ashtekar
variables. Where did the inspiration of these new variables come
from? Was Amitabha Sen's work a major influence?

``For me, personally, a lot of the inspiration came from results on
chiral solutions to the Einstein equations by Newman, Penrose and
Plebanski in the mid seventies. In a certain sense, this sector of
Einstein's theory is completely integrable. Therefore, it seemed
natural to try to base quantum theory on chiral variables. Sen's work
was definitely an important influence up to a certain point. About
the same time that Sen published his papers, Horowitz and I also had
some results on positivity of the Hamiltonian of GR (for open
universes) which used similar techniques. The two together suggested
the direction that I finally took."

What physical picture can you give us of them? You perform a
canonical transformation of the old geometrodynamical variables,
obtaining a connection and a triad as the new variables. A complex,
group-valued, self-dual, spin connection is hard to visualise. In what
way does a triad of orthonormal vectors on the 3-slice take over the
role of the extrinsic curvature?

``The connection enables us to parallel transport chiral fermions
along closed loops and therefore has a direct physical meaning.
Roughly speaking, the connection knows both about the (spatial
derivatives) of the triad and the extrinsic curvature (the
time-derivative of the triad). If you think of the 3-metric as the
analog of the position variable $q$ in quantum mechanics and extrinsic
curvature as the analog of $p$, then the connection is analogous to
the complex variable $z= q-ip$, which one often uses to construct the
so-called coherent states in quantum mechanics. Clearly, if you know
$(q, p)$ you know $(z, q)$ and vice versa. One can do quantum mechanics
with $(z, q)$ as basic variables, although it is somewhat unconventional.
In quantum GR, we are led to connections and triads because these are
the variables that simplify the equations."

The actual canonical transformation is non-linear. The result
casts GR as a dynamical theory of connections. The phase space of GR
is now seen to be embedded in that of Yang-Mills theory (a gauge
theory). The constraints form surfaces which define restricted motions
on this phase space. This hidden relation seems remarkable. Were you
expecting such a relation, or did it just drop out of the working?

``I knew that if the triads and connections can be shown to be
canonically conjugate, the embedding you refer to would be a
consequence. But it was far from obvious to me that the Poisson
brackets between triads and connections would be so simple. Also,
because of some subtle differences between the connections that Sen
was using and the ones I was led to use, it was not obvious to me that
all of Einstein's equations would be low order polynomials in the
triads and connections."

The next step is the  the transition from classical GR to quantum GR.
By using these Ashtekar variables, you are trying to quantise a theory of
connections, instead of a geometrodynamic theory based on spatial
3-metrics. In terms of connections, the constraint equations of GR
bear a great resemblance to those of {\it non-Abelian} Yang-Mills theory.

``Progress came about because there already existed a rich
machinery to deal with quantum gauge theories. We did have to make
some important changes eventually because in Yang-Mills theories, one
makes a heavy use of the background Minkowski metric and our framework
had to exist without reference to any background structures. But a number
of key idea came from Yang-Mills theories.

``Incidentally, because all other basic interactions can be formulated
in terms of connections, the geometrodynamic formulation did create a
distance between GR and other gauge theories. For example, Weinberg,
in the introduction to his book\footnote[10]{Gravitation and Cosmology:
principles and applications of the general theory of relativity,
Steven Weinberg,
New York, Wiley, 1972, xxviii, 657 p., ill, 23 cm.}
emphasises this point. However, both Einstein and Schr\"odinger had
presented a formulation of general relativity with connections, rather
than metrics as the basic variables. The reason the idea did not catch
on, I believe, is that their equations were even more complicated than
those of geometrodynamics. They used affine connections and
simplifications occur with chiral ones. I became aware of this piece
of history only recently."

\vspace{3mm}
{\Large{\bf Loop Variables}}
\vspace{3mm}

In QM we have operators and states. Operators act on states, and
take them from one allowed state to another. The simplest example of a
QM system may be the simple harmonic oscillator. There, the raising
and lowering operators take a particle subject to a constant restoring
force, to different energy eigenstates. The canonical variables,
position and momentum, are raised to the status of operators, and they
have to satisfy commutation relations. What are the corresponding
operators and quantum states in this approach?

``This is a somewhat technical point and I hope you will bear with
me.  In particle mechanics, we have the notion of configuration space
- the space of positions, for example - and quantum states are
functions on this space. In field theories, the situation is more
complicated. States are now functions on a {\it quantum} configuration
space which is a genuine enlargement of the classical configuration
space and the measures which dictate the inner product are typically
concentrated on the exotic, non-classical configurations.

``The precise domain space of quantum states is a well-defined
completion of the space of gauge equivalence classes of
connections. This space can be constructed without reference to a
background metric and we have also developed integral and differential
calculus on this space to define Hilbert spaces and operators. There is
no background metric or connection anywhere. This calculus tells us
that there is a well-defined `non-linear duality' between
connections and loops and hence one can also represent states as
suitable functions of loops. The connection and the loop
representations are `dual' to one another in the same sense as the
position and the momentum representations in quantum mechanics are
dual. In practice, issues related to the Planck regime are often more
transparent in the loop representation while semi-classical questions are
generally easier to analyse in the connection representation.

``For technical reasons connected with diffeomorphism invariance,
quantum states are represented by objects which are more general
than functions in the connection representation while in the loop
representation, one can regard them just as functions of loops.
In either case the basic quantum operators are associated with loops and
strips, i.e. ribbons. Their commutator algebra is surprisingly simple;
commutators can be expressed in terms of re-routing of loops and
strips, and intersections of loops with strips, and strips with strips.
It's a pretty, geometrical picture."

In the loop representation, quantum states arise as `functionals' of
Wilson loops. A deep relation exists between the equivalence classes of loops
and the theory of knot classes, invariants, and polynomials.
There is also a relation to Chern-Simons theory.

``This comes about because there is a specific `non-linear
duality' between loops and connections: Given a closed loop and a
connection, one obtains a number, the value of the trace of the
holonomy of the connection around the loop. It turns out that this
duality implies that there is a 1-1 correspondence between certain
functions on the loop space and measures on the quantum configuration
space. Therefore, we can take these loop functions as quantum states.
One can then formulate the diffeomorphism constraint rigourously and
seek its solutions. Not surprisingly, in the loop representation, the
solutions are (certain types of) knot invariants. Thus, after you have
imposed the diffeomorphism constraint, you can say that the quantum
configuration space is the space of knot classes and all knot
invariants/polynomials are the potential quantum states. At a
heuristic level, one expects the Chern-Simons theory to define a
diffeomorphism invariant measure on the space of connections and these
ideas have in fact been exploited by Gambini and Pullin to obtain
solutions to the Hamiltonian constraint. Whether such results can be
made rigorous is an open question; one would have to wait and see if
the current work by mathematicians in this domain leads to useful
results."

Are you feeding in what
the microstructure of spacetime should look like beforehand, or does
it come out of the equations? I have heard the analogy of Planck
spacetime being like a mesh of mail armour, made up of loops weaved
together.

``A key point in our approach is that we do not start out assuming
what quantum geometry should be like but let the theory lead us to
it. (Of course, the framework as a whole makes certain assumptions
- there should be no background structure and the Wilson loop
operators should be well-defined - but they are of a general nature).
We construct operators
that correspond to geometry and ask for states which, when probed with
these operators, would approximate a classical geometry at scales much
larger than the Planck scale. We then obtain specific states which
have this property but which display a discrete structure of a
definite type at the Planck scale. These are the states that look like
3-dimensional chain mail. The gravitational field is excited only at
the loops in the chains. However, on coarse graining, the states are
indistinguishable from classical geometries. More importantly, we find
that spectra of basic operators such as the area of a given
2-dimensional surface are quantised in the units of the Planck area.

``In the loop representation, all operators act by breaking, re-routing
and gluing of loops. These operations will code the quantum gravity
interactions. So, in the Planck regime, the familiar Feynman diagrams
will be replaced by `topological' diagrams representing these
operations. Feynman diagrams will arise only in a low energy regime
where, as I indicated above, the concept of gravitons is meaningful."

What was Profs. Smolin and Rovelli's inspiration to come up with
the loop representation? Has there been any collaboration with the
Trias and Gambini group, which is working on gauge theories?

``As I recall, they were trying to solve the diffeomorphism
constraint and realized that the solution is `obvious' in the loop
representation. At that time, we did not have the mathematical
machinery involving measures on the quantum configuration space and it
was hard to see how to construct solutions to this constraint in the
connection representation. Therefore, their realization was a real
breakthrough. Curiously, none of us knew about the work by Gambini and
his collaborators at that time. One of the students in our group,
Br\"{u}gmann, came upon their papers while applying the `loopy ideas'
to Yang-Mills theory and told us about them. We then contacted Gambini
and Trias. Since then, Gambini, Pullin and others have made some very
significant contributions. We are all extremely happy with these
synergetic developments."

You can also bring in supersymmetric matter couplings.

``Yes, supersymmetry can be brought in. There is work in this
direction by Jacobson, Matschull, Nicolai and others. The work by
Nicolai and his colleagues on 3-dimensional supergravity has been
especially illuminating; it showed that supergravity does allow
an infinite number of physical states rather than just, say, a
single solution analogous to the Hartle-Hawking wave function."

\vspace{3mm}
{\Large{\bf Current Work in the Program}}
\vspace{3mm}

Let us talk about some of the current research problems in the
field. In the language of connections, you have to complexify the
phase space of GR, since you have mixed (both complex and real)
variables. Now you have to restrict the phase space to real portions,
since the physical observables must be real quantities, not complex ones.
In other words, you must implement reality conditions in the quantum theory.
Have
these been dealt with satisfactorily?

``Several special cases had been treated satisfactorily over the
last few years. These include certain Bianchi models and midi
superspaces with a single Killing field. Over the past year, we have
been able to treat the {\it general case}. In the harmonic oscillator
case, the Segal-Bargmann transform enables us to pass from the
Schr\"odinger representation where the wave functions $\Psi(q)$ are
complex valued functions of $q$, to the holomorphic representation
where they are {\it holomorphic}\footnote[11]{A single-valued function $f(z)$
is
holomorphic (or regular or analytic) in a region if it is differentiable at
each
point of the region.} functions $\Psi(z)$ of $z$. Since we
know that $q$ and $p$ are self-adjoint in the Schr\"odinger
representation and since the Segal-Bargmann transform is unitary, you
know automatically that the reality conditions also hold in the
holomorphic representation; you don't have to check it explicitly. We
have constructed a rigorous analog of this transform for quantum
gravity. So, in principle, the problem has now been solved. However,
we still have to develop calculational tricks to make the transform
useful in practice."

One of the main problems has been to define an inner product
(vital to make sense of probabilities) in the Hilbert space. What is
the status of this search?

``This problem arises at two levels: kinematical and dynamical. One
generally needs an inner product on the the `kinematical' states
prior to imposition of constraints to make sure that the constraint
operators are well-defined and that there are no anomalies in the
constraint algebra. This part is taken care of by the recent
mathematical developments. Furthermore, as I indicated above, the
diffeomorphism constraint has been regularised and solved. There are
no anomalies. However, at the rigorous level, we still do not have a
complete treatment of the Hamiltonian constraint. Very recently, the
corresponding operator was constructed by Jerzy Lewandowski in Warsaw
but we are yet to examine the question of anomalies. If, there are
none, we can construct a complete set of physical states. Fortunately,
by now there are well-developed strategies to find the physically
appropriate inner product and the question is if one of them will be
easily implemented.  However, it is quite possible that to ensure
that there are no anomalies, we would have to bring in framed loops
and quantum groups and/or supersymmetry. Mathematical machinery does
exist to do this.  And I would find it quite exciting if, e.g., it is
the dynamics of quantum gravity that usher quantum groups into
physics.

``At a heuristic level, there has been a lot of progress on the issue of
solutions to the Hamiltonian constraint and Rovelli and Smolin have
introduced approximation techniques to find the inner product. These
are useful guidelines for rigorous work. However, the problem of inner
product on physical states is yet to be faced head-on.

``Incidentally, if the program is successful, we will have a quantum
theory of gravity at a level of rigour that exceeds the current level
in four dimensional quantum field theories. This may seem like an
overkill. However, since we have so little experience in
non-perturbative QFTs particularly in absence of a background
space-time, this level of care is needed both to ensure that we have
not ignored some important subtleties and to convince the skeptics
that a solution has really been found."

How is the problem of time evolution handled? If there are only
Hamiltonian constraints, and no Hamiltonian, how do you define a time
variable, and evolve the quantum system?

``Two approaches have been followed in both of which the notion of
time is an approximate one. The first is some work I completed in '89
where we saw that, in the connection representation, one can make a
certain truncation of the theory in which the quantum Hamiltonian
constraint can be written as the Schr\"odinger equation, where one of
the connection components serves as time. This shows how the familiar
evolution of QFT will arise in the low energy regime. The second and
more recent approach is due to Rovelli and Smolin. They couple gravity
to a scalar field and use the scalar field as time. So, while we still
don't know how to extract time and evolution in the Planck regime
- and, like many of my colleagues, I feel this is not necessarily a
meaningful thing to ask - in suitable approximations, time evolution has
been recovered."

For posterity, can I have your feelings as to where loop variables
are heading?

``There are three parts to the answer. The first refers to quantum
gravity the second to other physical theories of connection,
especially QCD, and the third to certain branches of mathematics.

``Within gravity, as I indicated above, for the first time, we have a
candidate for the quantum configuration space and the diffeomorphism
constraint has been formulated without anomalies, and we know its
general solution. If we can do the same for the Hamiltonian constraint
- possibly by bringing in framed loops and quantum groups - one could
say that quantum GR exists non-perturbatively. It is like saying, in
QCD, that we have a representation in which the Hamiltonian is
rigourously self-adjoint. This would be striking but not immediately
solve problems of physical interest. The situation would be the same
in QGR. However, knowing that the theory exists, one could then look
for reliable approximation methods and know that this is a meaningful
program. Many such methods are already being developed. I believe that
it is crucial to continue to look for such methods.

``Since the mathematical machinery that has been developed works for any
theory of connections, it is tempting to apply it to QCD. We have
already done so in two space-time dimensions and obtained several new
results. The most notable among these are closed expressions for the
Wilson loop Schwinger functions and a proof of equivalence of the
Hamiltonian and Euclidean path integral methods. One might venture
further and try to extend these methods to three and four space-time
dimensions. That they will lead to mathematically interesting quantum
field theories is clear. Whether these theories will be as useful in
physics is an open issue. In any case, it is exciting that one and the
same methods are being applied to theories of all basic forces of
Nature.

``On the mathematical side, there have also been some interesting new
results. A few years ago, it was widely believed that spaces of gauge
equivalent connections would not admit {\it any} non-trivial
diffeomorphism invariant measures. Several of us have found infinite
families of such measures by now. The relation between knots and these
diffeomorphism invariant measures is equally exciting. Our methods
also enable one to extend (sufficiently nice) invariants of ordinary
knots to those of generalised knots where loops are allowed to
intersect and can have kinks and overlaps. This is of interest to knot
and graph theorists. Such results arise because we are looking at the
familiar structure from a new angle, with a new perspective. Such
contributions, I believe, will continue."

\vspace{3mm}
{\Large{\bf Superstrings}}
\vspace{3mm}

Perhaps the most popularised of all the recent approaches to QG is
superstrings.

Superstring theory proposes that the elementary constituents of matter are
one dimensional curves, rather than point particles. According to the theory,
the quarks, leptons, and gauge bosons of the standard model (SM) arise as
excitation states of a truly fundamental entity - the superstring. These
excitations come from rotational or vibrational degrees of freedom (rather
like the harmonic modes in a violin string), or from internal degrees of
freedom like supersymmetry (SUSY) and Lie group symmetries. If they exist,
superstrings would be the smallest things in nature. They would be about
$10^{-33}$ cm long, and reside in the weird world of Planck scale physics.
Superstring excitations energies are about $10^{18}$ GeV, or $10^{-5}$ gm.
These particle masses are way too big
to be detected by current accelerators, including the Large Hadron Collider
(LHC) being built in Europe. We are looking at detecting particles as massive
as bacteria!

Superstrings can come in two varieties. One resembles a line segment with free
ends. This type is called the open superstring. The other type is a loop with
the topology of a circle - the closed superstring. Both versions reside in a
ten dimensional $(9+1)$ spacetime.

Since we only see four $(3+1)$ dimensions in real life, it is assumed that
the six extra spatial dimensions are curled up, in a region as small as the
Planck scale. The process of curling up is called compactification. There is
no mechanism in the theory to tell how you curl the extra dimensions up, and
this arbitrariness is a cause for concern. Theorists have explored Calabi-Yau
spaces, group manifolds, and other spaces as candidates. The idea of
compactification is reminiscent of Kaluza-Klein theory, which attempts
to unify gravity with electromagnetism using an extended spacetime of $5$ or
$(4+1)$ dimensions. This theory was extensively expanded in the 1970s
to include Yang-Mills theory which give, for example, the type of non-Abelian
gauge theory used in weak interactions (this involves a spacetime of
dimension greater than 5; e.g. 12 for SU(3) internal symmetry). The
mathematical
techniques of these larger-dimensional spacetimes (including what is
known as `dimensional reduction' - the means whereby one gets back
to the physical dimension 4) have been much used in discussions of
supergravity and superstring theory. However, it is important to
emphasise that these developments all work at the level of the
{\em classical} field equations: the quantum theory used in quantising
them is essentially standard.

String theories have an extraordinarily rich mathematical structure. Exploring
their symmetries, described by various algebras, has been a major focus of
work in the field. It has emerged that the same mathematical structures appear
in many different fields. For example, as it moves through spacetime, the
string traces out a world sheet or cylinder. Like the world lines (geodesics)
of point particles, this world surface is constrained to be of extremal area.
The symmetry properties of these surfaces are similar to those encountered
in two-dimensional condensed-matter systems. There has been valuable
cross-fertilisation between these fields.

Superstring theory is also related to subjects like 2D conformal field
theories, non-linear sigma models, WZW models, and other integrable models.
It has associations with 2D supergravity, Lie superalgebras, and W-algebras,
the latter being world-sheet symmetry algebras of bosonic string theories.
Quantization leads to problems with `anomalies' in these algebras whereby
classical symmetries are violated at the quantum level. Work is still being
done on anomalies in W-algebras and also more generally in Lie algebras.

The vibrational spectrum of open superstrings includes a massless spin-1 gauge
boson associated with a Yang-Mills group - $SO(32)$ or $E_8 \times E_8$. These
groups are needed for anomaly cancellations. For the closed superstring, a
massless spin-2 particle arises. It has been identified as the graviton, the
conjectured carrier of the quantum gravitational force.

GR is taken to be the large distance or low energy limit to the superstring.
Strings
seem like point particles at weak energy scales $(10^{-16} cm)$ since it is
many
orders of magnitude less than the superstring scale $(10^{-33} cm)$. A small
extended object looks like a point from the perspective of lower energies.

By examining string interactions, it is found that the existence of open
strings implies closed strings. Closed strings alone can form a consistent
system. It follows that every consistent string theory necessarily includes
gravity, since closed strings have gravitons in their excitation spectrum.

People have tried to generalise strings to higher dimensional objects, like
P-branes.
However it is very hard to make quantum theories of extended objects, and
get them to satisfy physical consistency constraints like
causality and unitarity. In fact, the more extended you go, the harder
it is. The successes of string theory rely on conformal symmetry, which is a
property solely of 2D world sheets.

The self-consistency of superstrings is heavily dependent on two
experimentally
undetected phenomena: supersymmetry and higher spacetime dimensions.

Ed Witten is at the Princeton Institute of Advanced Studies, New Jersey.
He is a major contributor to superstring theory.
He was once a history major at Brandeis, but had some science
background. His interest in superstrings was sparked by a review article by
John Schwarz\footnote[12]{Superstring Theory, Phys.Rept.89:223,1982.}. He says:
``Schwarz's review article made it easier to start learning about developments
in superstring theory than it had been before - but my concentrating
on reading it was, of course, a result of the fact that I was already
very curious about superstrings."

Quantum field theory began when Planck tried to explain the emission
spectrum of black body radiation. He postulated that energy was emitted
discontinuously in discrete units, or quanta. Later, from a statistical
analysis of the radiation field, Einstein suggested it wasn't just the
emission and absorption of radiation which was quantised, but the actual
electromagnetic field itself - in the form of photons. This idea was confirmed
experimentally by the Compton effect. Naturally, there were attempts to extend
the quantum field idea to the classical gravity field, described by a metric
tensor in Einstein's GR. The basic excitation of the quantum gravity field
would
be spin-2 radiation, which results from the symmetric nature of the metric
tensor.
How hard would it be to detect
the gravitational analog of the photon - the graviton, given the weakness of
the gravitational interaction?
Ed replies: ``Classical gravitational waves (for which there is indirect
evidence
from timing measurements of the binary pulsar) could be directly detected
by laser interferometers -- perhaps eventually giving also an important
new window into some exotic phenomena in astrophysics.  It is also possible
that a possible very long wavelength cosmic background of gravitational
waves could be detected by timing measurements involving space ships and
pulsars.  These experiments could indeed possibly verify directly the
spin two nature of gravitational waves.

``Unfortunately, there is no experiment in sight that could directly
verify the quantised nature of the gravitational field."

The gravity field is different to the other force fields since it plays a
dual role. One one hand, it defines the arena in which other fields interact.
Yet, gravity itself is a participant in the interaction. Does QFT assume that
spacetime is
flat to all energy scales i.e. has a Minkowskian spacetime signature even
at Planck scales?
``QFT as we know it, without quantising gravity, works on a fixed
space-time manifold, which for most obvious
purposes we can take to be Minkowski space. There is then a continuum
space-time down to arbitrarily small distances."

Gravity is highly non-linear since the gravity field couples to mass-energy.
The
gravitational field gravitates. Is this a major conceptual subtlety when we
try to quantise gravity?
``There are some conceptual issues in quantising gravity, but it is
not at all clear that they are decisive. What to me is the decisive
difficulty
is the problem of the infinities that arise when one tries to quantise
the usual theory." (It is instructive to contrast this point of view
with Abhay's, stated earlier on.)

\vspace{3mm}
{\Large{\bf Renormalisation, Infinities, and Anomalies}}
\vspace{3mm}

QFT works with perturbation expansions around a Minkowskian background.
Terms in these expansions are represented diagrammatically by Feynman diagrams.
The Feynman diagram has an attractive interpretation
as a representation of an actual physical process, and (though limited
to perturbation theory) this picture has some validity.

In QED interactions between photons and electrically charged particles are
linear. However the situation is different with gravitons. Since energy
gravitates, two gravitons
can bend each others path. Gravitons can also emit and absorb other gravitons
and couple to the other gauge bosons.

QFT introduces the concept of a particle's self-energy.
Let's take the electron self-energy as an example. According to QFT there is
a difference between a bare electron and a physical electron. A bare electron
is an unphysical entity because it has no radiation field around it. Such a
radiation field is known to exist because of Heisenberg's uncertainty
principle
which allows for the presence of a sheath of virtual photons. When a bare
electron interacts with this virtual field, the bare electron is converted
to a physical electron. This self-interaction affects the energy of the bare
electron, and hence its mass. Quantities like charge are affected as well. In
fact, all physical quantities are affected by the self-interaction.

When the Feynman integrals are evaluated for these self-energy terms, the
integrals diverge to infinity. A similar phenomena happen when we calculate
the photon self-energy in the form of vacuum polarisation diagrams.

The first order terms (tree level) are finite, but the higher order terms
(the radiative corrections) involving closed loops diverge. Physically,
the difficulties arise as a result of the point particle
concept. Heisenberg's principle says that energy conservation can be violated
locally such that the relation $\Delta E \cdot \Delta t \geq \hbar$ is obeyed.
Since an electron is taken to be a point, emission and reabsorption of a
virtual
photon need take no time at all. So the virtual photon can have an infinite
amount of energy. This contributes an infinite mass to the physical electron,
since it must haul around the virtual photon sheath. Other properties of the
electron affected similarly.

Despite these shortcomings, QFT is one of the most successful theories we have.
On this
Ed says: ``Just because we write down a quantum field theory Lagrangian
containing
some fields does not mean that we know what the particles are going to be.
QCD is an example where the particles (mesons and baryons) do not even
appear in the Lagrangian (which contains quarks and gluons).
Even if the particles are in natural one-to-one correspondence with the fields
- as in QED - the masses of particles cannot be just read off from the
constants in the Lagrangian.  It is necessary to compute all kinds of quantum
corrections. The fact that the nature and masses of the particles is
- in large part - computed, not postulated, is a manifestation of the fact
that the QFT description of nature is more thorough than its predecessors."

To handle these divergences, QFT invokes the principle that energy cannot
be measured, only differences in energy can be. A scheme called renormalisation
was developed by Feynman, Tomonaga, and Schwinger back in 1947 to take care of
this, for QED.

To handle divergent graphs, we firstly impose cut-offs on the integrals.
This modifies the theory into another one.
The cut-offs set some energy scale limit to the theory's predictions.
Ed comments: ``The cutoff theory has physically unacceptable properties, so it
is necessary to remove the cutoff and `renormalise.' This is just as well,
since the cutoff is arbitrary and if one did not have to remove it one
would lost most of the predictive power. In good cases, like QED, after
renormalisation the parameters in the quantum theory are in one-to-one
correspondence with the classical parameters; thus in QED the only parameters
are the charge and mass of the electron (the mass can be scaled to one
by a convenient choice of units)."

The second step is to renormalise the theory by incorporating the blanket of
virtual particles around the bare particle. This converts the bare particle
into a physical one. In the process, relations between bare and physical
quantities are introduced. These relations contain renormalisation constants.
Thirdly we revert from the regularised and renormalised theory to the QFT by
taking the limit when the cut-offs go to infinity. This restores the original
theory. The divergences reappear again, but only in the renormalisation
constants relating physical and bare quantities. Like the bare quantities, the
constants aren't observable. The original integrals, written in terms of
physical quantities, become well-defined and finite to all energy scales!
So in a sense we have shifted all the infinity problems aside.
According to Ed: ``Renormalization is natural and necessary even if the effects
are
finite. So there is no issue of not having to renormalise.
That would be so even if we were dealing with a fundamental theory
valid to arbitrarily short distances.

``In practice we are always (except maybe in string theory) dealing with
theories only valid down to very short distances, so there really is
a cutoff - we just do not know what it is. Renormalization then
extracts whatever can be learned without knowing what the cutoff is."

It would seem that you have to apply this procedure to all the terms in
the S-matrix expansion, for all the physical quantities you want to compute.
Now the expansion is an infinite series of terms, which seems to require
an infinite number of regularisation and renormalisation operations. Amazingly
QED has a gauge invariance which allows you to remove all the divergences from
the measurable quantities to all orders of perturbation expansion, via just
one type of renormalisation.

Why can't we compute physical quantities directly, instead of going via
this roundabout? Feynman himself thought renormalisation was `sweeping the
problem under the carpet'. He thought that we should be able to formulate
a theory such that we get zero ground state energy from the very start. What
physics could we be missing here? One possibility, already mentioned by Abhay,
is that the infinities ultimately
arise from neglecting effects in Planck scale physics which would have a
subtle effect on low energy physics.

When we try to fit gravity into QFT,
the algorithm is to linearise the GR field by choosing a co-ordinate system
in which the metric tensor can be expressed as $g_{\mu\nu} = \eta_{\mu\nu} +
h_{\mu\nu}$ where $\eta_{\mu\nu}$ is the Minkowski metric and $h_{\mu\nu}$ is a
small perturbation such that $|h_{\mu\nu}| \ll 1$. This small perturbation -
a tensor field, is subject to quantisation on a Minkowski background which
provides the causal structure. It is treated as the `graviton' particle.

``Formally there is no problem in constructing the perturbation expansion
of the S matrix along the above lines. The only real difficulty
is that one meets ultra-violet divergences that cannot be renormalised
away (without introducing new parameters absent in the classical theory)."

At the one-loop level pure QG's divergences can be renormalised away. However
the
divergences arising at the 2-loop level cannot be removed without robbing the
theory of predictive yield.

On the associated problems of anomalies in QG he
comments: ``The existence of a quantum theory is always delicate.
Once one writes down a classical Lagrangian, one does not automatically
get a quantum theory. Infinities and anomalies are two of the main ways
to fail. Anomalies that affect gauge invariance
are even worse than lack of renormalisability; they
mean that the perturbation expansion of the quantum theory just does
not make sense." Essentially the time rate of change of some quantity is zero
in the classical theory and upon quantisation, this rate becomes non-zero,
breaking
some symmetry, and hence a conservation law.

\vspace{3mm}
{\Large{\bf Recent Work in Superstrings}}
\vspace{3mm}

How does extending the idea of particles to superstrings change things?
An analogy would be helpful here. Let us take the decay of
the free neutron, which is mediated by the weak force. We know that the old
Fermi
theory of weak decay is not renormalisable. However, the electroweak theory is.
The electroweak theory replaces the point interaction of the old Fermi theory
by inserting a massive $W$ boson propagator, which mediates the decay. The
theory is now
renormalisable. Virtual supermassive superstring excitation states affect the
renormalisability of superstring interactions by `softening' or spreading out
the interactions.

Ed says: ``Certainly, in going to string theory, everything becomes softened,
and eventually finite. How best to understand it is not yet clear,
but somehow, the ultraviolet region is missing in string theory."

Although the superstrings are extended objects, their interactions still take
place at single spacetime points. However this point depends on the Lorentz
frame of the observer (the pants diagram phenomena) unlike the point particle
case. There, the spacetime interaction point is identical in all Lorentz
observer frames. Thus there less freedom in constructing string theories
compared to the point particle case. This is why we have only a finite number
of superstring
theories instead of the infinite number of point field theories. (We have to
choose the correct point-particle QFT via phenomenology.)

According to one counting scheme, there are currently three $d=10$ superstring
theories (Type I, Type II,
heterotic) and although they have no adjustable dimensionless parameters,
there are many mathematically consistent solutions which seem to satisfy the
superstring equations of motion. Choosing physically representative solutions
phenomenologically seems no different to inserting SM parameters by hand.
The problem of finding physical solutions can only be solved if we
know how compactification works. Doing this we can
find the ground (vacuum) state of the theory. From there, we can work out the
low energy excitation spectrum, find the ratios of the excitation masses, and
determine couplings constants from the topology of the compactified space.

Ed is optimistic about this arbitrariness: ``The attractive near-uniqueness
that
string theory appears
to possess in principle is largely lost in practice because one finds
a plethora of possible classical solutions.  Hopefully, in time we will
learn something new - perhaps connected with the question of why the
cosmological constant vanishes - and the nature of the issue will change."

On the problem of formulating a non-perturbative theory of superstrings Ed
says: ``My work with Nathan Seiberg (and related work he did by himself) led to
some
new results about non-perturbative behaviour of some quantum field theories,
but not yet to new results on non-perturbative behaviour of string theory.

String theory appears to admit a group of discrete field transformations,
called $S$-dualities, as exact non-perturbative quantum symmetries. How could
these symmetries help in formulating a non-perturbative QG theory of strings?

``I don't know yet, but I think it is a very exciting clue.
$S$-dualities are symmetries that have no real analog in our previous
experience in physics (they do persist in some field theories
- including some of those I studied with Seiberg - which are related
to low energy limits of string theories)."

Particles trace out world lines, but strings trace out world sheets or
cylinders, which can be treated as Riemann surfaces.
When the Feynman diagrams are worked out for the closed string, it is found
they finite order by order in the genus expansion.

Ed says: ``In string theory everything is finite - there is no need for
renormalisation.

``Feynman diagrams are replaced by Riemann surfaces; these have
much additional symmetry and structure, and this is related to some
of the known wonders of the subject - there are surely also unknown
wonders ($S$-duality is a hint of them), but since they are unknown I cannot
tell you about them."

On the matter of progress on the physical principles behind superstrings
which should generalise Einstein's equivalence principle, he replies:
``The new physical principles haven't yet been found.  It is hard to say
how much progress there has been, since some of the things that have
been learned may well turn out to be important clues."

Superstrings are supposed to reside in 10 spacetime dimensions, but we are
really only speaking approximately. The phase space of the superstring consists
of all the possible orientations and states the string can be in.
What are the problems of formulating superstrings directly in
four spacetime dimensions from the start? Aren't superstrings supposed
to generate spacetime, instead of just residing in it, since superstrings have
gravitons
in their excitation spectrum?

``It is true that in our present understanding 10 spacetime dimensions
is the classical solution of string theory with the maximum symmetry.
Exactly how strings can be so smart as to generate space-time is still
a mystery, but they do.

``You can formulate strings `directly in four dimensions from the start'
but then you do not have the degree of uniqueness that one has in 10
dimensions. That is why the theory is often said to be 10 dimensional."

There is no mechanism in superstring theory which tells how us the extra
dimensions should compactify. Feynman said about superstrings, ``I don't like
it
that they're not calculating anything. I don't like it that they don't check
their ideas. I don't like it that for anything that disagrees with experiment,
they cook up an explanation - a fix-up to say `Well, it still might be true.'
For example the [superstring] theory requires 10 dimensions. Well, maybe
there's
a way of wrapping up six of the dimensions. Yes, that's possible mathematically
but why not seven? When they write their equation, the {\em equation} should
decide how many of these things get wrapped up, {\em not} the desire to agree
with experiment..." What is your response to this remark?

``Some of the predictions are indeed checked. For instance, strings
predict gravity while pre-string physics makes quantum gravity impossible.
This prediction has been tested - and the result is one of the main
reasons for the interest in string theory."

Are you familiar with the Ashtekar canonical gravity program? If you are,
do you think that the program is founded on valid physical assumptions?

``Some of the work in that program is interesting to me.
I'd personally be surprised, however, if it turns out that conventional
general relativity exists as a non-perturbative quantum field theory."

Do you think the LHC will shed any light on low energy superstring
phenomenology? The detection of supersymmetry partners could be crucial.
To explain the absence of SUSY partners
at low energies we have introduced the idea of the spontaneous breaking of
supersymmetry, in analog to the Higgs mechanism for electroweak symmetry
breaking. One facet of the SM which is unconfirmed is this mysterious Higgs
field
which pervades all spacetime, which makes the electroweak theory work by giving
masses to particles, and which we have no shred of evidence for.

``Apart from gravity, supersymmetry is one of the general predictions
of string theory. Therefore, confirming it at the LHC would give
a big boost to the whole effort. Moreover, if supersymmetry is observed,
it won't be just that experimentalists will say `Yes, there is
supersymmetry.'
They will observe a plethora of superpartner masses and couplings.
There is a reasonable hope that in all that data there will be clues
to how string theory should be done - or tests of theoretical ideas
that may have emerged in the meantime."

I asked Ed what the inspiration for his ideas was. Did he
manipulate page-long equations in his head? Did he deal with concepts, and then
phrase everything in maths later on? Where Einstein, Galileo, and Newton were
different from everyone else was they could ask really simple questions which
had
shattering consequences for existing physical theories. Was this Ed's modus
operandi too?

To this he replied: ``I am afraid I don't work on the cosmic scale
you are suggesting.
Most of the time I am bogged down doing nothing, and whatever successes
I have had come by focussing on a very little bit of the total picture."

\vspace{3mm}
{\Large{\bf Outlook}}
\vspace{3mm}

It is clear from the discussion that we are a long way off from even the
foundations of a full theory of QG. Despite some famous claims, the end of
theoretical physics is still not in sight. I asked Chris about possible lines
of future investigation.

``I think that the superstring programme and the Ashtekar programme are
currently the only two really viable approaches to trying to construct a full
quantum theory of gravity, and I therefore support strongly work on both
fronts!

``Myself, I have long been bothered by the use of continuum ideas in
both quantum theory and classical general relativity, whereas both
programmes do take this for granted. So I would like to be able to
get away from this, but the problem is to find a theoretical
framework that does not seen to contrived and artificial. One of the
many reasons why the superstring and Ashtekar programmes {\em are} so
important is that they do not have this contrived nature: within
their own terms each has quite compelling reasons for taking it
seriously.

``Similarly, I (like Roger Penrose) believe in my heart that quantum
theory itself needs to be changed radically to reflect our changing
conception of space and time. But, again, it is difficult to do this
in a way that is believable to one other person than the author of
the paper (Salam's definition of what it means to say that a theory
is worth taking seriously!)"

Ed Witten gives this advice to those who want a try at superstrings:
``It is hard to give general advice to young researchers. But people who want
to learn strings should definitely make sure that along the way they get a
thorough knowledge of what is known in particle physics and how it can be
(more or less) derived from strings."

As to where superstrings are heading he has this to say:
``Lots of really nice things have been discovered, and that will continue.
Sooner or later, probably far in the next century, the underlying
unifying ideas will be discovered and then the physical landscape will
be very different from what we can imagine today.  If we are lucky and work
hard, maybe we will live to see the day!"

\vspace{3mm}
{\Large{\bf Acknowledgements}}
\vspace{3mm}

This article was based on interviews with Chris Isham, Ed Witten and
Abhay Ashtekar. William Spence at Queen Mary and Westfield College in London
helped review the final draft. It was
the basis for a theory seminar presented at Melbourne University, Australia.

\vspace{3mm}
{\Large{\bf Further Reading}}
\vspace{3mm}

{\bf The Problem of Time}

\blankline

{\em Prima Facie Questions in Quantum Gravity}, C.J. Isham,
Canonical Relativity: Classical and Quantum, eds J. Ehlers
and H. Friedrich, Springer-Verlag, Berlin (1994).

\vspace{2mm}

{\em Conceptual and Geometrical Problems in Quantum Gravity}, C.J. Isham in
``Recent Developments of Quantum Fields'', pp123--225, eds
H. Mitter and H. Gaustener, Springer-Verlag (1992).

\vspace{2mm}

{\em Canonical Quantum Gravity and the question of Time}, C.J. Isham in
``Relativity, Classical and Quantum'', eds J. Ehlers and H. Friedrich,
Springer-Verlag (1994), e-Print Archive: gr-qc/9310031.

\vspace{2mm}

{\em Canonical Quantum Gravity and the problem of Time}, C.J. Isham in
``Integrable Systems, Quantum Groups, and Quantum Field Theories'',
pp157--288, eds L. A. Ibort and M. A. Rodriguez,
Kluwer Academic Publishers, London (1993).

\vspace{2mm}

{\em Continuous Histories and the History Group in Generalized Quantum
Theory}, C.J. Isham and N. Linden, Imperial -TP-94-95-19, Mar 1995, 24pp,
e-Print Archive: gr-qc/9503063.

\blankline

{\bf The Ashtekar Program}

\blankline

{\em Bibliography of Publications related to Classical and Quantum Gravity
in terms of the Ashtekar Variables}, B. Br\"{u}gmann (Syracuse U.),
SU-GP-BIB, Mar 1993. 14pp. pre-Print Archive: gr-qc@xxx.lanl.gov - 9303015.

\vspace{2mm}

{\em Loop Representations}, B. Br\"{u}gmann (Munich, Max Planck Inst.),
MPI-PH-93-94, Dec 1993, 38pp. In ``Bad Honnef 1993, Proceedings, Canonical
gravity", pp213-253, and Muenchen MPI Phys. - MPI-Ph-93-094 (93/12,rec.Dec.)
38 pp. pre-Print Archive: gr-qc/9312001a.

\vspace{2mm}

{\em Canonical Quantum Gravity}, Karel V. Kuchar (Utah U.),
UU-REL-92-12-10, n.d. (recd Apr 1993) 40pp.  e-Print Archive:
gr-qc@xxx.lanl.gov - 9304012.

\vspace{2mm}

{\em Recent Mathematical Developments in Quantum General Relativity},
Abhay Ashtekar (Penn State U.), CGPG-94-9-4, n.d. (recd Oct 1994)
13pp. e-Print Archive: gr-qc@xxx.lanl.gov - 9411055

\vspace{2mm}

{\em Overview and Outlook}, Abhay Ashtekar (Penn State U.), in
Canonical Gravity from Classical to Quantum, edited by J.Ehlers and
H. Friedrich, Springer-Verlag (1994).

\vspace{2mm}

{\em Quantum Gravity: A Mathematical Physics Perspective}, Abhay
Ashtekar (Penn State U.), in The Proceedings of the International Conference
on Mathematical Physics Towards the 21st Century, Negev, Beer Sheva,
Israel, 14-19 Mar 1993; edited by R. N. Sen and Gersten (Beer-Sheva
University Press, 1994).

\vspace{2mm}

{\em Mathematical Problems of Nonperturbative Quantum General Relativity},
Abhay Ashtekar (Syracuse U.), SU-GP-92-11-2, Dec
1992. 87pp. Invited talk at Lectures given at Les Houches Summer
School on Gravitation and Quantisations, Session 57, Les Houches,
France, 5 Jul - 1 Aug 1992.  e-Print Archive: gr-qc@xxx.lanl.gov -
9302024

\vspace{2mm}

{\em Conceptual Problems of Quantum Gravity. Proceedings, Osgood Hill
Conference, North Andover, USA, May 15-18, 1988}, A. Ashtekar,
(ed.) (Syracuse U.), J. Stachel, (ed.) (Boston U.), 1991.  Boston,
USA: Birkhaeuser (1991) 602 p. (Einstein studies, 2).

\vspace{2mm}

{\em Lectures on Nonperturbative Canonical Gravity}, A. Ashtekar
(Syracuse U.), 1991.  Singapore, Singapore: World Scientific (1991)
334 p. (Advanced series in astrophysics and cosmology, 6).

\vspace{2mm}

{\em New Variables for Classical and Quantum Gravity}, A. Ashtekar
(Syracuse U. and UC, Santa Barbara), 1986.  Phys. Rev. Lett. 57 ( 1986)
2244-2247.

\vspace{2mm}

{\em Experimental Signatures of Quantum Gravity}, Lee Smolin (Penn State
and Princeton, Inst. Advanced Study). CGPG-95-3-2, Mar 1995.
e-Print Archive: gr-qc/9503027

\vspace{2mm}

{\em Spin Networks and Quantum Gravity}, Carlo Rovelli (Pittsburgh U.),
Lee Smolin (Penn State U. and Princeton, Inst. Advanced Study).
CGPG-95-4-4, Apr 1995. 42pp. e-Print Archive: gr-qc/9505006

\vspace{2mm}

{\em Discreteness of Area and Volume in Quantum Gravity}, Carlo Rovelli
(Pittsburgh. U.), Lee Smolin (Penn State U.). CGPG-94-11-1, Nov 1994. 36pp.
e-Print Archive: gr-qc/9411005

\vspace{2mm}

{\em Quantization of Diffeomorphism Invariant Theories of Connections
with Local Degrees of Freedom}, Abhay Ashtekar, Jerzy Lewandowski
(Warsaw U., ITP), Donald Marolf (UC, Santa Barbara), Jose Mourao
(Algarve U.), Thomas Thiemann (Penn State U.). UCSBTH-95-7, Apr 1995. 71pp.
e-Print Archive: gr-qc/9504018.

\vspace{2mm}

{\em Differential Geometry on the Space of Connections via Graphs and
Projective Limits}, Abhay Ashtekar (Penn State U. and Newton Inst. Math. Sci.,
Cambridge), Jerzy Lewandowski (Warsaw U., ITP and Schrodinger Inst., Vienna).
CGPG-94-12-04, Dec 1994. 53pp. e-Print Archive: hep-th/9412073.

\vspace{2mm}

{\em Coherent State Transforms for Spaces of Connections}, Abhay Ashtekar,
Jerzy Lewandowski, Donald Marolf, Jose Mourao, Thomas Thiemann.
CGPG-94-12-02, Dec 1994. 38pp. e-Print Archive: gr-qc/9412014

\blankline

{\bf Superstrings}

\blankline

{\em A One Loop Test of String Duality}, Cumrun Vafa (Harvard U.), Edward
Witten (Princeton, Inst. Advanced Study). HUTP-95-A015, May 1995, 13pp.
e-Print Archive: hep-th/9505053.

\vspace{2mm}

{\em Classical Duality Symmetries in Two Dimensions}, John H. Schwarz
(Cal Tech). CALT-68-1994, May 1995. 13pp. Presented at STRINGS 95:
Future Perspectives in String Theory, Los Angeles, CA, 13-18 Mar 1995.
e-Print Archive: hep-th/9505170.

\vspace{2mm}

{\em String Theory Symmetries}, John.H. Schwarz. CALT-68-1984,
Mar 1995. 12pp. Presented at International Workshop on Planck Scale
Physics, Puri, India, 12-21 Dec 1994. e-Print Archive: hep-th/9503127.

\vspace{2mm}

{\em Evidence for Nonperturbative String Symmetries}, John H. Schwarz.
CALT-68-1965, Jul 1994. 9pp. Conference on Topology, Strings and Integrable
Models (Satellite Colloquium to the ICMP-11 18-23 Jul 1994), Paris, France,
25-28 Jul 1994. e-Print Archive: hep-th/9411178.

\blankline

\vspace{6mm}
\hrule

\end{document}